# NaMemo2: Facilitating Teacher-Student Interaction with Theory-Based Design and Student Autonomy Consideration

Guang Jiang, School of Telecommunications Engineering, Xidian University, Xi'an, China, gjiang@mail.xidian.edu.cn

Jiahui Zhu, School of Telecommunications Engineering, Xidian University, Xi'an, China, 20011210497@stu.xidian.edu.cn

Yunsong Li, School of Telecommunications Engineering, Xidian University, Xi'an, China, ysli@mail.xidian.edu.cn

Pengcheng An, School of Design, Southern University of Science and Technology, Shenzhen, China, anpc@sustech.edu.cn

Yunlong Wang, Institute of High Performance Computing (IHPC),
Agency for Science, Technology and Research (A*STAR), Singapore,
wang_yunlong@ihpc.a-star.edu.sg

**Abstract**: Teacher-student interaction (TSI) is essential for learning efficiency and harmonious teacher-student interpersonal relationships. However, studies on TSI support tools often focus on teacher needs while neglecting student needs and autonomy. To enhance both lecturer competence in delivering interpersonal interaction and student autonomy in TSI, we developed NaMemo2, a novel augmented-reality system that allows students to express their willingness to TSI and displays student information to teachers during lectures. The design and evaluation process follows a new framework, STUDIER, which can facilitate the development of theory-based ethnics-aware TSI support tools in general. The quantitative results of our four-week field study with four classes in a university suggested that NaMemo2 can improve 1) TSI in the classroom from both teacher and student perspectives, 2) student attitudes and willingness to TSI, and 3) student attitudes to the deployment of NaMemo2. The qualitative feedback from students and teachers indicated that improving TSI may be responsible for improved attention in students and a better classroom atmosphere during lectures.

---

**Introduction**

Research on *smart classroom* technologies has sought to empower teachers to better understand and improve teaching-learning activities (Saini & Goel, 2019), e.g., tracking student learning states (Ahuja et al., 2019; Holstein, McLaren, & Aleven, 2018) and enhancing teacher interaction with students in the classroom (An et al., 2019; Saquib, Bose, George, & Kamvar, 2018). Studies also focused on online virtual classes to tackle challenges such as maintaining student engagement (Dumford & Miller, 2018). The teacher-student interaction (TSI) is a fundamental aspect of pedagogy that influences student behavioral engagement (Nguyen, Cannata, & Miller, 2018), autonomous motivation of learning (Opdenakker, Maulana, & den Brok, 2012), resilience (Liebenberg et al., 2016), and learning achievement (Pianta, 2016; Sun et al., 2022). In this paper, we focus on technology design and development to support TSI in university classrooms. Although there have been studies exploring the use of TSI support tools in university classrooms (Ahuja et al., 2019; Alavi & Dillenbourg, 2012; Alavi, Dillenbourg, & Kaplan, 2009; Fernandez-Nieto, An, Zhao, Buckingham Shum, & Martinez-Maldonado, 2022), there remain many unanswered research questions and technical needs from teachers and students for improving TSI in the university context.

TSI primarily takes the forms of physical proximity, conversation, body language, eye contact, or communication through shared (digital) media. Appropriate TSI contributes to efficient learning and a positive climate in the classroom and harmonious teacher-student interpersonal relationships (Hagenauer & Volet, 2014; Li & Yang, 2021; Sun et al., 2022). A classic yet effective way for teachers to bond with students is to address them by their names in teaching practice (Cooper, Haney, Krieg, & Brownell, 2017; Glenz, 2014; Tanner, 2011, 2013). However, this method has become challenging for university teachers, given the large number of students in a class. A teacher in a university typically only remembers the names of a few students for the entire duration of a course, which affords these students better chances for interactive participation in the course and subconsciously missed personal engagement with many of the rest students. On the other hand, student needs, willingness, and autonomy for interpersonal interaction with teachers during lectures have not been explicitly conveyed to the teachers, which may also hinder teacher efforts to interact with students. These challenges and demands have been highlighted by researchers in recent years (Cooper et al., 2017; Tanner, 2013; Wubbels & Brekelmans, 2005). Therefore, there is a need for novel tools to support teachers in recalling student information (e.g., names) and student expression of personal needs (e.g., the willingness toward and preference for interacting with teachers during lectures) for better TSI.

We found a theory-based design framework for the development of TSI support tools to be absent in the literature. Without integrating theories into the design and evaluation process, it is difficult to accumulate knowledge and identify opportunities in the research domain. Therefore, in addition to developing a TSI support tool, we also explore a holistic approach to guide the development of TSI support tools, which leads to our proposal of the new framework, STUDIER (i.e., Sparking, Targeting & Understanding, Designing & Implementing, Evaluating & Refining). In STUDIER, we integrate theoretical models of TSI (Pianta & Hamre, 2009) and behavioral theories (Michie, van Stralen, & West, 2011a) to guide the design of TSI support tools towards discipline progress, and avoid research that merely focus on improving user experience (Long, 2021).

We also emphasize the consideration of ethics and accessibility principles to frame the design to be accountable and responsible.

Following the STUDIER framework, we developed an augmented-reality TSI support tool, NaMemo2, that allows students to convey their willingness to interact to the teacher as well as show their names to the teacher in physical classrooms. This information can help teachers reduce their cognitive load of having to remember many names and increase their awareness of student needs and willingness to interact during lectures. As the system uses facial recognition technology, we carefully considered ethics in our use case and integrated student control of facial photos and interaction willingness into the system. Our system, NaMemo2, was developed by redesigning the hardware platform and software pipeline of its predecessor, NaMemo (Jiang et al., 2020). Following the privacy and autonomy principles of facial recognition technology use (Doberstein, Charbonneau, Morin, & Despatie, 2022; Smith & Miller, 2022), we designed new features for student control and autonomy to explore the ethical application of facial recognition in schools (Andrejevic & Selwyn, 2020). Using the system UI, once the teacher attempts to interact with a student, they can obtain each student's name and willingness to interact using the NaMemo2 user interface (Figure 2). The results of our intervention study suggest that our method can improve both student attitudes to TSI and the deployment of NaMemo2 in the classroom.

In the subsequent sections, we will first review related work in the literature. Next, we describe the development of NaMemo2 using the STUDIER framework. We next report on the evaluation of NaMemo2, followed by discussions of our evaluation results and the lessons we learned through our study.

**Related Work**

To illustrate the foundations of and challenges in developing TSI support tools, we review research on the theories of TSI, frameworks for designing TSI support tools, and existing TSI support tools.

***Theories of Teacher-Student Interaction***

Prior empirical studies showed that TSI can influence student behavioral engagement (Nguyen et al., 2018), autonomous motivation in learning (Opdenakker et al., 2012), resilience (Liebenberg et al., 2016), and learning achievement (Pianta, 2016; Sun et al., 2022). Drawing on theories and empirical studies (Brophy, 1999; J. S. Eccles & R. W. Roeser, 1999; Pressley et al., 2003), Hamre and Pianta proposed the *Teaching through Interactions* (TTI) model that describe three domains of classroom teacher-student interaction: emotional support, classroom organization, and instructional support (Hamre et al., 2013; Pianta & Hamre, 2009). Each domain includes multiple dimensions (e.g., positive climate and teacher sensitivity for emotional support), and each dimension is described by explicit indicators (e.g., positive communications and respect for a positive climate). Each indicator is eventually linked to specific behaviors (e.g., physical proximity and peer assistance) that can be observed and measured. The TTI model presents a Domain-Dimension-Indicator-Behavior structure to conceptualize and operationalize teacher-student interaction. Following the TTI model, researchers and practitioners can design experiments to investigate how specific interactions contribute to teaching outcomes or realize appropriate strategies to optimize certain interactions. With a similar concept (teacher-student contact), Korthagen et al. examined the opinions of teachers and students on timing, characteristics, and the meaning of good teacher-student contact (Korthagen,

Attema-Noordewier, & Zwart, 2014). The TTI model and related empirical studies provide a solid foundation that can inspire and frame the design and development of technologies to properly support TTI. Meanwhile, translating the theories and their implications to real-world technologies for improving TSI remains challenging.

In another research line of Dynamic Assessment (DA) (Poehner & Wang, 2021), rooted in Sociocultural Theory (Vygotsky, 1978) and the Mediated Learning Experience framework (Feuerstein, Rand, Hoffman, & Vig, 1979), researchers developed methods that allow teachers to mediate students' performance and understand their learning potential (Poehner & Wang, 2021; Swanson & Lussier, 2001). Between the two major types of DA (interactionist and interventionist), interactionist methods emphasize dialogic interaction between teachers and students to improve student performance, which is more malleable and challenging to teachers (Ghonsooly & Hassanzadeh, 2019). However, technologies to support teachers in identifying opportune moments for interaction during DA are underexplored. Our work can also benefit this domain by informing the design of tools for interactionist DA.

### Frameworks for Developing TSI Support Tools

To facilitate the co-design of learning analytics systems with stakeholders, Holstein et al. summarized a five-step process: initial needs analysis and concept generation, initial concept validation, iterative lower-fidelity prototyping, iterative higher-fidelity prototyping, and iterative classroom piloting and experimental evaluation (Holstein, McLaren, & Aleven, 2019). From the scope of teaching augmentation, the TA framework frames the design space into five dimensions, namely target (teacher abilities), attentional level, social visibility level, presence over time, and interpretation (An, Holstein, D'Anjou,

Eggen, & Bakker, 2020). These design frameworks provide practical tools to design and analyze smart classroom technologies but lack direct connections to the theories and principles of teacher-student interaction. Integrating and elaborating on existing theories can help researchers to accumulate knowledge and identify opportunities in this research domain. For instance, TSI studies can incorporate the taxonomy in the TTI model to avoid using different terms for the same concepts when reporting the study results. On the other hand, developing technologies for a specific domain ideally requires the cooperation of interdisciplinary teams of designers, developers, end-users, and domain experts. Therefore, holistic design frameworks are necessary to synthesize considerations and best practices from different disciplines to guide the development of TSI support tools. In this paper, we propose a new design framework for developing such TSI support tools and demonstrate how it has been applied in the development of a new TSI support tool.

### *TSI Support Tools and Student Autonomy in TSI*

Along with the development of information and communication technologies in the last few decades, we have seen many digital tools being developed and applied to school classrooms to support teacher-student interaction and other teaching practices (Harper, 2018; Saini & Goel, 2019). However, our literature review revealed most existing tools to focus on sensing TSI, e.g., applying distributed sensor networks (Saquib et al., 2018), indoor positioning (Martinez-Maldonado et al., 2020), and using computer vision technology (Ahuja et al., 2019) to track teacher proximity and student behavior in the classroom. The tracked data can provide real-time *feedback* to teachers on their attention and time distribution for the students in the classroom (An, Bakker, Ordanovski, Taconis, & Eggen, 2018; An et al., 2019; Saquib et al., 2018), student attention and emotion (Saquib et al., 2018), and student learning states (Holstein et al., 2018). We identified a gap between sensing technologies and the use of student autonomy for TSI support in previous studies. Existing TSI support tools emphasize the role of the teachers in the interaction while neglecting the student expression of their need and autonomy. In this study, we therefore develop a new TSI support tool to show student names and TSI willingness to teachers during lectures, which helps teachers to overcome the difficulty of remembering many names and increases their awareness of student needs and willingness to interact during lectures.

## Methodology: Developing NaMemo2 with STUDIER

In this section, we present our new design framework for developing TSI support tools and illustrate how we applied it to develop a new TSI support tool.

A design framework is not a theoretical model illustrating relations between factors; instead, it is a guideline that highlights key steps and design elements to help practitioners plan, design, develop, and evaluate a target product. Design frameworks are well established in many domains, e.g., health behavior change (Michie et al., 2011a; Mohr, Schueller, Montague, Burns, & Rashidi, 2014; Mummah, Robinson, King, Gardner, & Sutton, 2016), user-centric explainable AI (Miller, 2019; D. Wang, Yang, Abdul, & Lim, 2019), and collective intelligence and creativity (Cox, Wang, Abdul, Weth, & Lim, 2021; Malone, Laubacher, & Dellarocas, 2009). These design frameworks are especially useful for interdisciplinary and emerging research fields, where the design process involves multi-aspect factors and considerations.

The development process of NaMemo2 consists of *Sparking, Targeting & Understanding, Designing & Implementing,* and *Evaluating & Refining* (STUDIER), which we conceptualize into a design framework. Inspired by frameworks and reviews in other domains, e.g., *Creativity Support Tools* (Frich, Biskjaer, & Dalsgaard, 2018) and *Behavior Change Support Tools* (Oinas-Kukkonen, 2013; Y. Wang, Fadhil, Lange, & Reiterer, 2019), this framework emphasizes the explicit integration of relevant theories and design principles to encourage holistic thinking, and theory-driven and ethics-aware design. In this section, we illustrate each part of STUDIER (Figure 1) and demonstrate the design process for NaMemo2.

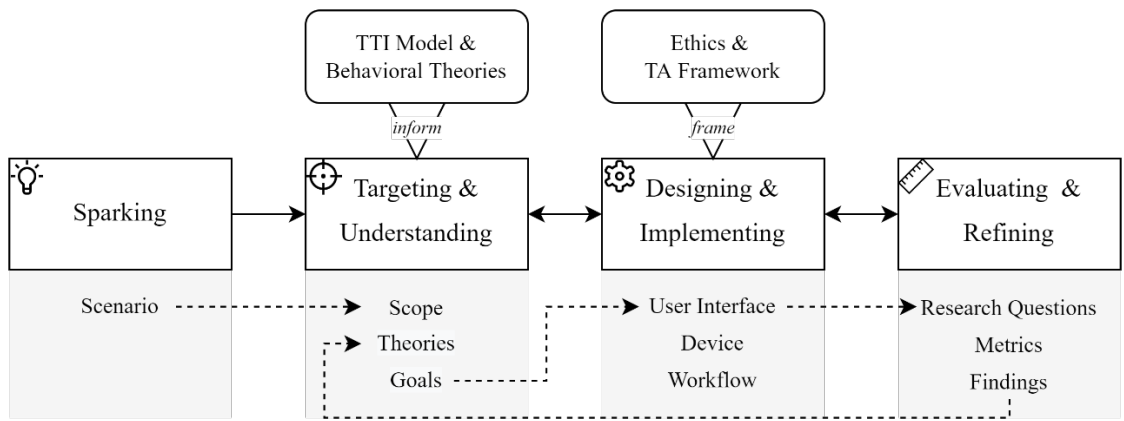

**Figure 1**: The diagram of STUDIER. The arrows with solid lines connect the major steps. The arrows with dotted lines show the connections between strongly relevant elements across the major steps.

### *Sparking*

The initial problem may be identified by teachers, students, or other stakeholders of the education ecosystem in their daily practice of teaching, learning, or organizing teaching-learning activities. We use *spark* to indicate the process of realizing a pain point and initiating an intention to solve the problem in TSI. In this stage, one needs to describe the *scenario* of the identified problem as richly as possible with contextual information.

In the sparking stage of NaMemo2, one author (Teacher T) found it challenging to remember many student names when interacting with students during lectures. As he described:

> *"I teach an engineering subject in a university. I believe that teachers interacting with students in class is essential for inquiring about students' learning states, drawing students' attention, and improving the classroom atmosphere. However, I find it inefficient and sometimes frustrating to address students when I don't know their names in class. It's quite challenging to remember so many students' names with two lectures per week. I need to point to the student with my hand, count the row and column numbers of the seat, or describe the cloth color or hairstyle of the student. All these methods take more time than I expect and break the class fluency. In addition, using the pointing gesture or describing students' appearance is not polite and respectful to students."*

### Targeting & Understanding

Given the described scenario, we next need to define the scope of the target *user group* and *interaction problem*, which will help us articulate the overarching *goal*. This *targeting* process complements the process of *understanding* the problem context and characteristics of the user group. In Teacher T's scenario, the target user group is intuitively the university teachers, because they have limited interaction time to be familiar with the students. K-12 teachers may not have such difficulty because they have more time for teaching and only need to pay attention to small groups of students. To confirm the generalization of the proposed problem, the design team discussed the described scenario with five of Teacher T's colleagues, who all agreed on the significance of the problem in the discussion.

The primary need of the teachers is to address students by their names during lectures in the classroom. Meanwhile, we should not ignore the need and autonomy of students in our design. Given that TSI is reciprocal by nature, and that student needs for TSI can be highly personal, it would be meaningful to allow student input for their preferences and willingness for TSI over time. Therefore, we need to develop a system that allows the teachers to know student names in real time, at the same time allowing the students to express their needs and willingness for interaction, which serves as the *design goal*. In addition, we also need an *intervention goal* that defines the expected effect of the target users using our system. We refer to the TTI model as a taxonomy to understand and conceptualize the potential effect of resolving the target interaction problem (Pianta & Hamre, 2009). For Teacher T's case of knowing student names, we found two relevant dimensions: 1) *positive climate* through improved respect and communication, and 2) *better productivity* through efficient transitions. Therefore, we should examine if the system could increase the quantity and quality of teacher-student interaction in empirical studies.

Informed by *behavioral theories* (Michie, van Stralen, & West, 2011b), the behavior of the teachers in interacting with students should be driven by their motivation of TSI. Therefore, we should clarify that we design the system to support teachers who have the motivation for TSI, instead of those that need to be motivated. While we design the tool for teachers as the primary users, the students are the passive users who need to be considered and respected. Student attitudes to system deployment, as well as the change in their attitudes to TSI after using the system, should thus be important measures for system evaluation.

### Designing & Implementing

Following the design and intervention goals, we then design the *user interface* (UI), select *devices*, and define the system *workflow*. We found that the most intuitive way to indicate student names in real-time is to use augmented reality (AR) technology based on facial recognition, which displays student information on top of the real-time captured panorama in the UI (Figure 2).

The facial recognition function requires students to upload photos of their faces taken from different angles. Even though all students had previously submitted their biometric face photos for their university student ID card, some of them may still hesitate to upload face photos for identification. We therefore studied and followed the recently discussed ethical principles of using facial recognition technology (Smith & Miller, 2022). First, we limited our use of student photos to the name indication function in NaMemo2. Second, we granted students full control of photo usage: Students uploaded their photos after informed consent and could withdraw their permission at any time. Furthermore, to support student expression of willingness to interact in class, we included the Willingness to Interact (WtoI) score, which is set by the students when showing their names. Eventually, the system workflow involves 1) students uploading photos and setting their willingness to interact before lectures, 2) teachers setting up the system in the classroom before each lecture, and 3) teachers interacting with the UI in class when requiring student names.

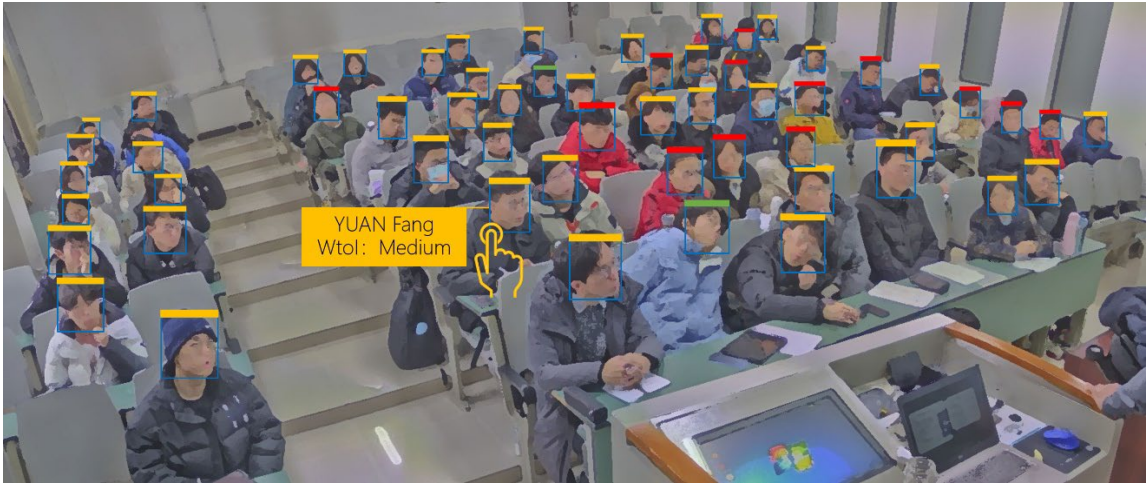

**Figure 2**: A showcase of NaMemo2 UI a teacher used in a class. The boxes locate the recognized student faces. The bar colors on top of the boxes show student willingness to interact (WtoI). Green means high, yellow means moderate, and red indicates low willingness. The student's name with WtoI will appear when the box surrounding their face is clicked. The image is obscured and an alias is used to protect student privacy, and the original Chinese text in the UI is replaced with English for demonstration.

Our system consists of three hardware parts: a pan-tilt camera capturing real-time images, a terminal (PCs, tablets, or smartphones) accessing the user interface, and an edge computing unit (ECU) running the backend server to connect the camera and user terminal. The pan-tilt camera is placed in the front area of the classroom to scan all the students and send the images to the server, while the server hosted in the ECU runs the face recognition algorithm and renders the UI. Using any terminal with a web browser, the teacher can access the UI and obtain student names and willingness to interact by clicking on the UI.

According to the design dimensions in the TA framework (An et al., 2020), we tend to build a low-attentional (for both teachers and students), low-social visibility (only teachers see the UI), and real-time system that accurately recognizes students in a large classroom and minimizes interruption to the teaching flow. We iterate our solutions by stepwise resolving the design requirements. (1) We could use a camera with a telephoto lens to maintain the resolution of faces in images so that students sitting in the last row can also be recognized. However, the telephoto lens cannot cover the whole classroom in one frame. (2) Therefore, we chose a PTZ camera, which can rotate the camera for a broader view. However, the camera's movement and affiliated noise may distract students. (3) To minimize the obtrusiveness, we limited the rotation speed to 10 degrees/second while pausing the camera when snapping to prevent motion-induced image blur. The reduced speed may cause a delay in the UI update, but the delay will not harm the ease of identifying students through the UI in normal classroom settings where students do not change positions in general during the lecture.

The key challenge is to accurately recognize multiple faces at different viewing angles since students may be looking at the teacher, blackboard, and book on the table during the lecture. To achieve a high recognition accuracy to avoid adverse impacts on the teacher experience, we developed an algorithm pipeline combining a few computer vision methods, including image stitching (Brown & Lowe, 2007), instance segmentation (He, Gkioxari, Dollar, & Girshick, 2017) to avoid stitching-caused artifacts, and face recognition (Deng, Guo, Ververas, Kotsia, & Zafeiriou, 2020).

### *Evaluating and Refining*

The final stage of our design framework is evaluating the effectiveness of the developed system and refining it when necessary. A series of feasibility tests (for system function testing), formative studies (for usability testing), and summative studies (for answering research questions) can be conducted. Upon major revision of the system functions and UI, the process can be iterated across stages to adjust the elements accordingly. Ideally, the findings from the evaluation results should be compared with existing theories and contribute to our understanding of TSI.

Before conducting a summative evaluation of the system, we conducted a feasibility study with one university teacher and 161 students in two classes for ten lectures, which we reported previously (Jiang et al., 2020). The primary aim was to collect enough data to test facial recognition in a large classroom with many students. We therefore selected a teacher participant who 1) had a good relationship with students and 2) actively interacted with students in class. The teacher helped us motivate the students to participate in the study, and eventually, all 161 students participated in the feasibility study. The facial recognition accuracy in our test was 99.2%. In the next section, we report on our follow-up study to explore the effects of NaMemo2 on TSI in the classroom, in addressing our intervention goal.

**Evaluation of NaMemo2**

We have illustrated the development process for NaMemo2 along with an illustration of the STUDIER framework. In the section "Targeting & Understanding", we defined the design goal (i.e., reminding teachers of student names and supporting students to express interaction needs and willingness) and the intervention goal (i.e., quantity and quality of TSI), which guided the realization of NaMemo2. In this section, we report on the summative evaluation of the interventional goal in our design.

We plan to answer three research questions in this evaluation study:

(1) Can NaMemo2 support teachers for TSI? We expect teachers to find NaMemo2 useful in reminding them of student names, to make classroom TSI more fluent and frequent.

(2) Will student attitudes to and willingness for TSI change after using NaMemo2 in their class? Student attitudes to TSI refer to their general assessment (positive vs. negative) of the TSI in a particular class, while willingness refers to their intention to get involved in TSI. Both factors have been investigated in public administration, e.g., payment for public safety (Donahue & Miller, 2006), recycling e-waste (Song, Wang, & Li, 2012), and participation in research studies (Trauth, Jewell, Ricci, Musa, & Siminoff, 2000). We adopt the two factors to reflect student perception of the potential increase in TSI during class.

(3) Will student attitudes to NaMemo2 change after using NaMemo2 in their class? Even though there are privacy protection strategies in NaMemo2 through data access control and limiting the use of facial recognition to name indication, some students may still show concerns about the use of facial recognition. We expect student attitudes to NaMemo2 to be improved after its deployment.

***Participants***

To answer our research questions, we deployed NaMemo2 in four classes for four weeks at a university in China during November and December 2021. The institutional ethics committee approved the study. We next advertised our study plan via the university mail list and the social media accounts of the authors. This recruitment method was convenient and commonly used for exploratory evaluation in HCI studies (e.g., (Luo et al., 2018; Y. Wang, König, & Reiterer, 2021; Y. Wang & Reiterer, 2019)). Six teachers contacted us for participation, and two of them were randomly selected for participation in this study due to the limited availability of our hardware. Each teacher taught two classes of the course "Micro-Controllers: Theory and Applications" to second-year undergraduate students.

***Study Procedure and Metrics***

The study began in the middle of the 2021 winter semester, so that we could apply a single-arm study design with pre-intervention measures and post-intervention measures. The pre-intervention measures served as the baseline for the post-intervention measures. Before field deployment, we introduced the workflow and UI of NaMemo2 to the teacher participants, who then helped us introduce the system to their students in class. Immediately after the introduction to students, we delivered a pre-study survey to the students in the four classes via their WeChat groups to obtain their informed consent, demographics, and baseline measurements of the metrics (Table 1). To help students think through the survey questions and avoid poorly thought out ratings, we designed open-ended questions to describe their interactive experience with the teachers during lectures and their rationale for their rating scores. We did not provide any incentive for the participants to avoid potentially confounding effects in their responses. The students volunteered to anonymously fill out the survey. Meanwhile, we posted the link in the WeChat group for students to upload at least three photographs of their faces (i.e., one taken from the front and two taken from the side) to the NaMemo2 system for the study. The students who uploaded photos needed to provide their actual names and willingness for interaction (High, Medium, Low) to be shown in the NaMemo2 UI (Figure 2).

During the four-week field deployment, the students were allowed to log in to the system to change their willingness level. Before each lecture, a teaching assistant set up the hardware in the classroom to allow the teacher to log in to the system with their device for access to NaMemo2 UI.

After the field deployment, we conducted post-study surveys for both students and teachers. In the student survey, we added the perceived change of TSI in class (see Table 1) to the metrics used in the pre-study survey. In the teacher survey, we measured the perceived helpfulness of NaMemo2 (on a 7-point Likert scale) and collected teacher feedback on suggestions to refine NaMemo2.

**Table 1**: The metrics in the pre-study and post-study student surveys. The survey questions were originally in Chinese. We have translated them into English in this table.

| Metric | Description | Survey |
|---|---|---|
| Attitudes to TSI | - What is your general attitude to TSI in class? [7-point Likert scale (-3 to 3): Very negative, Negative, Somewhat negative, Neither negative nor positive, Somewhat positive, Agree, Strongly positive]<br>- Why? [Free text] | Pre-Study & Post-Study |
| Willingness to TSI | -Do you hope to have more interaction with your teacher during the lecture? [7-point Likert scale (-3 to 3): Strongly disagree, Disagree, Somewhat disagree, Neither disagree nor agree, Somewhat agree, Agree, Strongly Agree]<br>-Why? [Free text] | Pre-Study & Post-Study |
| Attitudes to NaMemo2 | - What is your general attitude towards using NaMemo2 to support TSI in the classroom? [7-point Likert scale (-3 to 3): Very negative, Negative, Somewhat negative, Neither negative nor positive, Somewhat positive, Positive, Strongly positive]<br>- Why? [Free text] | Pre-Study & Post-Study |
| Perceived Change of TSI | - What is the change in TSI after your teacher used NaMemo2 in the classroom? [7-point Likert scale: Decreased a lot, Decreased a little, No change, Increased a little, Increased a lot]<br>- And any other changes during the lecture, e.g., class atmosphere? [Free text] | Post-Study |

### Data Collection and Analysis

A total of 118 students (71%) uploaded their photos into our system. We collected 105 and 126 valid entries in the pre-study and post-study student surveys, respectively. The participants were 21.9% female, between 18 and 22 years old (M=19.3). We applied both quantitative and qualitative methods to analyze the collected data. We used Pearson's *r* to evaluate correlations between variables, two-tailed t-test for the mean difference between groups, and Cohen's *d* for the effect size (Nakagawa & Cuthill, 2007). All statistical tests were conducted in JMP Pro 15.2.0 (SAS, 2022), which is a frequently used tool in scientific research (e.g., in (Cox et al., 2021; Kelkar, Boushey, & Okos, 2015; Y. Wang, Venkatesh, & Lim, 2022)). The significance level was set to 0.05. We are aware of the debate on using statistical methods to analyze Likert scale responses. and chose to use parametric tests as they are generally more robust than non-parametric tests (Fagerland, 2012; Sullivan & Artino, 2013).

For the qualitative analysis, we performed a thematic analysis to accumulate evidence to further explain the statistical results. We used open coding in grounded theory (Glaser & Strauss, 2006) to derive categories and affinity diagramming (Beyer & Holtzblatt, 1998) to consolidate categories to themes supporting the understanding and explanation of our research questions. Two authors coded the free-text answers in the surveys independently, then resolved the inconsistency among the coding themes by discussion.

***Results***

*(RQ1) Can NaMemo2 support teachers for TSI?*

We examined how the students perceived changes in TSI and what the teachers perceived of NaMemo2 after using it in order to answer this question. We found that 67% of the students surveyed regarded TSI to have *increased a little*, while 11% of the students perceived a *large increase* after the deployment of NaMemo2 (Figure 3). In the free-text responses of students to the question of "other changes in class," the dominant themes were *higher focus in class* (n = 82) and *livelier atmosphere* (n = 33). Some students additionally mentioned that *student attendance rate seemed higher* (n = 5) and *there were fewer students sitting in the back rows* (n = 3). But there remained a number of students who either felt that TSI did not change (n = 30) or felt that they did not pay attention if it changed (n = 9).

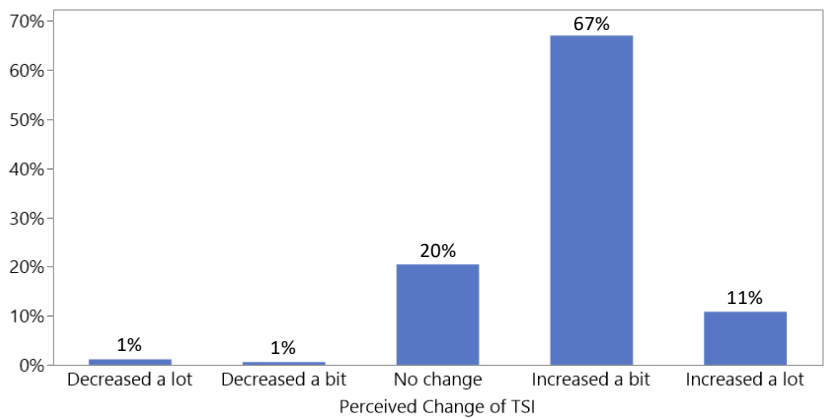

**Figure 3**: The rating distribution of student perceived change in TSI.

The two teachers consistently reported that they often used NaMemo2 during lectures and found the system helpful for interacting with students. We identified three main themes from the teacher feedback: *better focus and engagement of students*, *better* *classroom atmosphere*, and *higher teacher-student respect*. "*I felt my students paid more attention to my lecture, and the engagement improved the class atmosphere*," said T1 (Teacher Participant 1). T2 commented that "*I used the [NaMemo2] system to call on a student by name to start a conversation, which showed my respect to my students; I hope they could feel the respect and have a better impression of me.*"

*(RQ2) Will student attitudes and willingness to TSI change after using NaMemo2 in their class?*

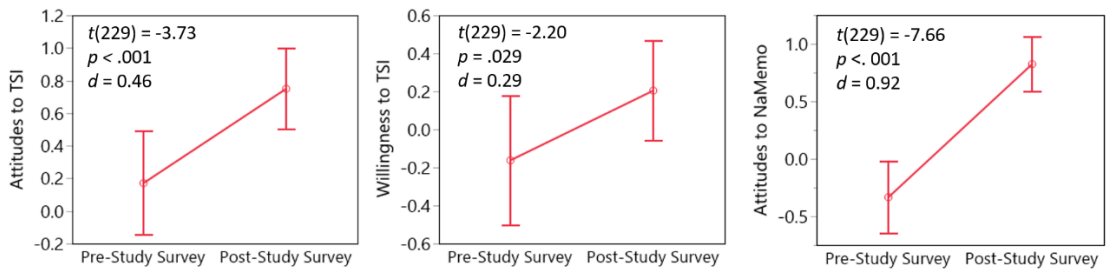

**Figure 4**: Results of student ratings on attitudes to TSI, willingness to TSI, and attitudes to NaMemo2 from the pre-study and post-study surveys.

The results showed a significant improvement in student attitudes toward TSI ($t(229)$ = -3.73, $p < .001$, $d = 0.46$) and willingness to engage in TSI ($t(229)$ = -2.20, $p = .029$, $d = 0.29$) over the course of the study (Figure 4). The key themes behind the positive attitudes and willingness to engage in TSI included *higher focus in class* (n = 80), *better learning efficiency* (n = 34), and *higher sense of participation* (n = 17). Student Participant 8 (S8) pointed out that "*even though I felt a bit nervous, but I know this could force me to be more focus in class,*" while S102 explained that "*I felt my teacher cares about us when he asked several students if I understand that part.*" These examples indicated a change in the classroom atmosphere, which might have improved student attitudes to TSI. We also found the increasing willingness to engage in TSI to be possibly due to the improvement in the classroom atmosphere. "*I didn't want to interact with my teacher before because I didn't follow the lecture; later when I became more focused, I understand more and become more willing to interact in class because I feel more confident,*" said S23. Some students even became more active to deliberately interact with the teacher, as S40 stated, "*my teacher did not call me quite often. Instead, I deliberately asked questions for clarification. Not sure if she knows my name, but I feel good interacting with my teacher in class. It will be better if she knows me.*" On the other hand, we found the main reasons for negative attitudes and low willingness to TSI include *worry of failing to answer questions* (n = 11), *shyness of speaking in class* (n = 7), and *nervousness* (n = 4).

From the correlation test, student attitudes to TSI are closely related to their willingness for TSI, $r = .558$, $p < .001$, $d = 0.92$, which is consistent with evidence in other domains of behavior change research, including participation in research studies (Trauth et al., 2000) and paying for public safety (Donahue & Miller, 2006).

*(RQ3) Will student attitudes to NaMemo2 change after using NaMemo2 in their class?*

We found a significant improvement of student attitudes to NaMemo2 over the course of the study, $t(229) = -7.66$, $p < .001$, (Figure 4). From the pre-study survey, we identified the main reasons for negative attitudes to NaMemo2 as *disliking facial recognition* (n = 8), *the perception that teachers knowing names would not be necessary for TSI* (n = 8), *disliking TSI* (n = 4), and *worry over the failure to answer questions* (n = 3). Some students reported a positive change in their attitude toward NaMemo2 after using it. As S55 commented, "*I don't see any harm when using face recognition in this case. I do feel the* *change after my teacher used this system that we became more focused and much less checking my phone. So my attitude to the system is more positive than before*." Importantly, students appreciated that teachers cared and knew their names. S17 additionally reported, "*I really appreciate it that the teacher cares about our names*," and S61 commented "*The system can help my teacher know my name anytime in class, so it's easier for my teacher to remember me. I think there are not many ways to improve teacher-student communication, but this could be a good way*."

These findings align with a study on the use of name tents to identify students to teachers, where students were more willing to use name tents after realizing the benefits of having their names known to teachers (Cooper et al., 2017).

### Summary and Implications for Refinement

The perceptions of both students and teachers suggest that NaMemo2 can support TSI by increasing both the quantity and quality of TSI. Comparing the results of the pre- and post-study surveys, we observed a significant improvement in student attitudes and willingness for TSI, as well as their receptiveness to the use of NaMemo2.

In response to the open-ended question about the expected improvement of NaMemo2 in the post-study survey, both teachers expressed a desire for more supportive functions for TSI. One participant pointed out the need and difficulty of knowing student learning status during lectures, which can help the teacher arrange the timing of interaction with students. "*… for example, if the system can remind me when the percentage of 'head-down' students is getting higher, I will slow down and interact more with students*." Another participant wanted to see the relationship between where the students sat and their willingness to interact, so that he could direct more attention to the less active "area" in the classroom. This participant also wanted to have reminders for interaction, because "*when the lecture is dense, I may forget to interact with students*". These suggestions for refining NaMemo2 could inform our follow-up studies.

We demonstrated each step in STUDIER along with the development of NaMemo2. As a design framework, STUDIER could also be used to guide the review of relevant studies of TSI support tools.

## Discussion

### *Perspectives from Behavioral Theories*

Our results suggest that the deployment of NaMemo2 can improve TSI during lectures by indicating student names for teachers on demand. We refer to behavioral theories to explain this effect. According to Fogg's behavior model (Fogg, 2009) and the Capability, Opportunity, and Motivation Model of Behavior (Michie et al., 2011b), motivation, ability, and triggers (opportunity) are the three components that generate a target behavior that in turn influences those components. When NaMemo2 reduces the difficulty of remembering many student names, which is a barrier for teachers, teachers became more likely to interact with students. The positive perception of TSI from students, as a reward, could increase teacher motivation for TSI. As suggested by the teacher participant feedback, future work may also consider providing context-aware timely reminders for TSI, because teachers might not necessarily realize opportunities to interact with students or forget to do so when focusing too much on the teaching material. The requirement of additional functions from teacher participants also implies the need for technical support for teachers regarding Dynamic Assessment methods. Besides the timely reminders, NaMemo2 can also enable post-lecture data analysis to identify students' status (e.g., "*head-down*" patterns) along the lecture slides, which may inform the teacher to add questions or pauses in the complex parts for students to understand better.

### *Encouraging Participation with Student Autonomy*

The positive outcome of deploying NaMemo2 would not be possible without carefully considering the privacy and autonomy of the students. This was not a problem in previous studies on TSI, since these were conducted in K-12 schools, where the student concerns about privacy and autonomy were not often obvious problems (Hafen et al., 2015; Hamre et al., 2013). However, the teacher-student relationship, student-student relationship, as well as classroom culture/atmosphere in university classrooms are all different from the K-12 setting. As such, we need to understand the context and address student concerns. In the case of NaMemo2, we understand the concerns of using facial recognition technology (Ahuja et al., 2019; Andrejevic & Selwyn, 2020; Doberstein et al., 2022). Even though we hope to apply this powerful tool to benefit students, we limit its use to only indicate names. Our study thus illustrates an exploration of ethical use of facial recognition technology in schools: 1) participants were informed about the usage of their data (i.e., only for name-indicating in NaMemo2); 2) we granted participants full control of the data collection (i.e., they were given the choice of whether their photos were uploaded or not); and 3) participant willingness and right (for TSI in our case) was well-respected even if they did not provide data for the system. As system developers, we hope for target users to accept and benefit from the system. Our ethics consideration provides an option for hesitant students to gradually learn and appreciate the system before they accept and participate in it.

Besides mitigating the privacy concerns of students, we observed that other groups of students who tend to be introverted (Kriegel, 2022) or face more difficulty in learning the course material than other students (Noble, 2022) also hesitated to participate, according to the qualitative analysis of student feedback. NaMemo2 thus provides these students with a platform to express their willingness to interact, which prompt them to think about the benefits of TSI and to observe the behavior of their peers during TSI. We feel that allowing the students to express themselves can promote the fair treatment of students in the classroom (Bernard E. Whitley, Jr. & Deborah Ware Balogh, Patricia Keith-Spiegel, 2000). Future studies are warranted to confirm the effect of NaMemo2 on classroom fairness for students.

### *Limitations and Generalization*

Despite positive results, our evaluation study suffers from a few limitations. Firstly, the sample size of participants is small, and the type of lecture studied is not diverse. Therefore, generalizing the findings to broader target users should be treated with caution. Secondly, the measure of TSI is based on participant recall and perception in the post-study survey. Despite the importance of user perception, solely using subjective metrics may be prone to cognitive bias and recall accuracy. Future studies should consider objective measures of TSI, which may need to integrate technologies like visual scene understanding and audio recognition into the system (Ahuja et al., 2019). Thirdly, the metrics of student attitudes to TSI and NaMemo2 are summary perceptions, while dividing attitudes into different aspects may lead to more comprehensive measures (Donahue & Miller, 2006). Lastly, the NaMemo2 platform includes additional hardware (i.e., a pan-tilt camera and edge computing unit) in traditional classrooms, which can be inconvenient to use.

While our evaluation covered only a small aspect of the broad domain of TSI, the demonstrated STUDIER framework could be applied to many other topics that are relevant within this domain. By explicitly including the TTI model (Hafen et al., 2015; Hamre et al., 2013; Pianta & Hamre, 2009), behavioral theories (Fogg, 2009; Michie et al., 2011b; Schwarzer, 2008; Y. Wang et al., 2019), and the TA framework (An et al., 2020) in the design process, we aim to better review, relate, and accumulate relevant studies of TSI support tools.

**Conclusion**

In this paper, we presented the development and evaluation of NaMemo2, an augmented-reality system to support teacher-student interaction (TSI) in the university classroom by 1) applying computer vision methods to indicate student names to teachers, and 2) empowering students to express their interaction willingness to teachers through a real-time interface. Through the design process of NaMemo2, this paper demonstrated STUDIER (Sparking, Targeting & Understanding, Designing & Implementing, Evaluating & Refining), a design framework for the domain of TSI support tools, emphasizing design in a theory-driven and ethics-aware manner towards principled progress in this domain. NaMemo2 granted students TSI autonomy in class, which provided an ethics-aware use case of facial recognition in the university setting. The evaluation results suggest that NaMemo2 can improve TSI quantity and quality, student attitudes and willingness to TSI, and their attitudes to technology deployment in classrooms.

**Acknowledgement**

Pengcheng An acknowledges funding support from Shenzhen Grant for Universities

Stability Support Program (Grant Number: 2022081517130800).